\documentclass[sn-mathphys]{sn-jnl}

\usepackage{amsmath,amssymb}
\usepackage{color}

\jyear{2023}%

\begin{document}

\title[Article Title]{Thermodynamic cost of Brownian computers in the stochastic thermodynamics of resetting}


\author*[1]{\fnm{Yasuhiro} \sur{Utsumi}}\email{utsumi@phen.mie-u.ac.jp}

\author[2]{\fnm{Dimitry} \sur{Golubev}}\email{dmitry.golubev@aalto.fi}
\equalcont{These authors contributed equally to this work.}

\author[3]{\fnm{Ferdinand} \sur{Peper}}\email{peper@nict.go.jp}
\equalcont{These authors contributed equally to this work.}

\affil*[1]{\orgdiv{Department of Physics Engineering, Faculty of Engineering}, \orgname{Mie University}, \orgaddress{\city{Tsu}, \postcode{514-8507}, \state{Mie}, \country{Japan}}}

\affil[2]{\orgdiv{QTF Centre of Excellence, Department of Applied Physics}, \orgname{Aalto University}, \orgaddress{\postcode{P.O. Box 15100}, \state{FI-00076 Aalto}, \country{Finland}}}

\affil[3]{\orgdiv{National Institute of Information and Communications Technology}, 
\orgaddress{\street{Iwaoka-588-2 Iwaokacho, Nishi Ward}, \city{Kobe}, \postcode{651-2492}, \state{Hyogo}, \country{Japan}}}


\abstract{We analyze the thermodynamic cost of a logically reversible Brownian Turing machine operating in the first-passage time protocol based on the stochastic thermodynamics of resetting. 
In this framework, the thermodynamic cost of computation is the reset entropy production, which is interpreted as the information reduction by a resetter external to the computer. 
At the level of a single trajectory, the reset entropy production is associated with unidirectional transitions and is a function of the time-dependent distribution probability. 
We analyze an approximation that replaces the distribution probability with the empirical sojourn time, which can be obtained at the single-trajectory level. 
The approximation is suitable for the numerical analysis by the Gillespie algorithm and provides a reasonable average value for the reset entropy. }




\maketitle

\section{Introduction}

The study of the thermodynamic cost of computation was initiated more than half a century ago during the early stages of modern computers~\cite{Landauer1961,Bennett1982,Bennett1985}. 
A notable example is the thermodynamic cost of information erasure: Landauer's bound~\cite{Landauer1961} states that the minimum energy dissipation in erasing one bit of memory at temperature $T$ is $k_{\rm B}T \ln 2$, where $k_{\rm B}$ is the Boltzmann constant.

Recently, this issue has been attracting renewed attention from the viewpoint of the recently established theory of information thermodynamics~\cite{Sagawa2019,Wolpert2019,Kolchinsky2020,Ito2013}. The modern technical tool behind these works is stochastic thermodynamics~\cite{Seifert2012,VandenBroeck2013,VandenBroeck2015}, which introduces thermodynamic quantities at a single stochastic trajectory level, such as stochastic entropy, energy, heat, and work. This view naturally applies to mesoscopic systems, such as solid-state quantum devices~\cite{Klages2012,Pekola2015}.
During the last two decades, stochastic thermodynamics has turned out to be a powerful tool to analyze information-related experiments, such as information-to-work conversion~\cite{Klages2012,Pekola2015}.

%
%

A standard theoretical model of the thermodynamics of computation is the Brownian computer \cite{Bennett1982}. 
It utilizes random thermal transitions for searching through a labyrinth of the configuration space isomorphic to the desired computation. 
The thermodynamic cost of a Brownian computer has been a matter of argument~\cite{Norton2013}. 
Brownian computation is argued to be thermodynamically irreversible from the analogy of one `Brownian particle' gas expanding irreversibly into a box, which is the analog of the state space $ \Omega $ of a computation model. 
During the computation, the system entropy increases due to the Brownian particle exploring the state space by $k_{\rm B} \ln   \vert \Omega \vert$, where $\vert \Omega   \vert$ is the size of the state space. 
The cost is interpreted as the amount of external work necessary to reset the computer to the initial zero entropy state in a computation cycle~\cite{Strasberg2015}.

%
%

The Brownian computer is designed to output a unique solution corresponding to a given input, although the computation time is not deterministic due to thermal fluctuations. 
Therefore, the computation time is important for characterizing its performance. 
However, previous arguments on the thermodynamic cost did not take it into account. 
In this paper, we consider the first-passage time as the computation time and analyze the cost of Brownian computation in the framework of the stochastic thermodynamics of resetting \cite{Fuchs2016}. 
This paper elaborates on our previous work on the thermodynamic uncertainty relation in a Brownian computer~\cite{utsumi2022computation}, especially focusing on the thermodynamic cost part.
{\color{black} The information cost and the first-passage time is a topic discussed recently~\cite{toledomarin2022passage}.}

The structure of this paper is as follows. 
In Section~\ref{LRBTM} we introduce the logically reversible Brownian Turing machine by focusing on its stochastic dynamics. 
Section~\ref{STR} reviews the first-passage time protocol and the stochastic thermodynamics of resetting. 
In Section~\ref{RD} we present our numerical results, and in Section~\ref{conclusion} we summarize the results.

\section{Stochastic Dynamics of Logically Reversible Browninan Turing Machine}\label{LRBTM}

\begin{figure}[ht]%
\centering
\includegraphics[width=1\textwidth]{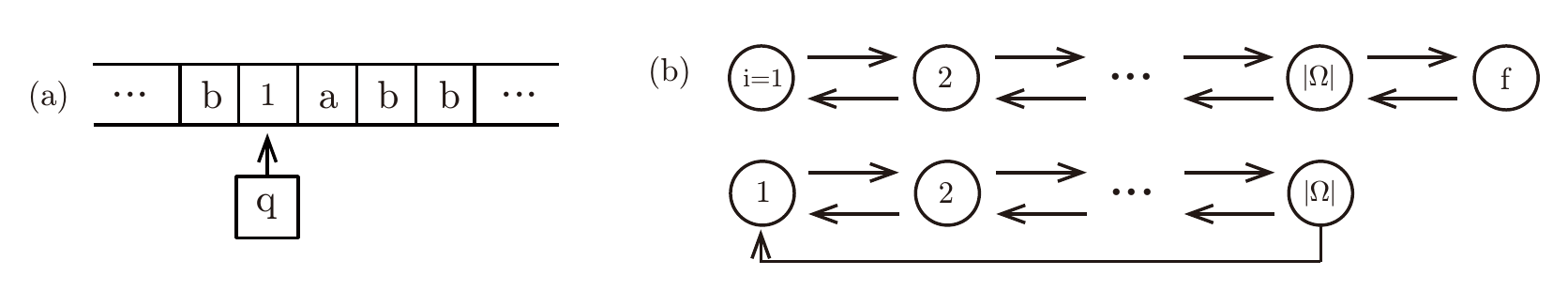}
\caption{
(a) Schematic picture of a Turing machine (TM). 
(b) The transition state diagrams of the logically reversible TM without reset (top panel) and with reset (bottom panel). 
}
\label{fig:rtm}
\end{figure}

Figure~\ref{fig:rtm} (a) is a schematic picture of a Turing machine (TM)~\cite{Bennett1973} consisting of a read/write head (arrow) plus a finite control (box) with a finite number of internal states (denoted by $q$) and a two-way infinite tape. 
The tape consists of an infinite sequence of tape cells. 
The TM updates a symbol on a cell, the state of the finite control and the position of the tape head according to a program expressed as a set of quintuples.
Such an update procedure is depicted by a transition state diagram, see top panel of Fig.~\ref{fig:rtm} (b). 
In the case of the logically reversible TM (RTM), there is a one-to-one correspondence between adjacent states~\cite{Bennett1973,Strasberg2015,utsumi2022computation}, whereas the size of the state space $\vert \Omega \vert$ varies depending on the program and its input. 
The RTM starts from the initial (start) state $\vert {\rm i} \rangle = \vert 1 \rangle$ and updates its state sequentially until it reaches the final (halt) state $\vert {\rm f} \rangle = \vert \vert \Omega \vert +1 \rangle$, which is the solution, as follows: $\vert 1 \rangle \to \vert 2 \rangle \to \cdots \to \vert \vert \Omega \vert \rangle \to \vert {\rm f} \rangle$.

The Brownian RTM is constructed by introducing stochastic transitions between adjacent states.
Then, the transition rate matrix of the Brownian RTM is simply that of a finite one-dimensional chain~\cite{Strasberg2015}:
\begin{align}
L_{\rm RTM} =& \sum_{\omega=2}^{\vert \Omega \vert} \left[ - (\gamma^+ + \gamma^-) \vert \omega \rangle \langle \omega \vert + \gamma^+ \vert \omega \rangle \langle \omega-1 \vert + \gamma^- \vert \omega \rangle \langle \omega+1 \vert \right] \nonumber \\
&- \gamma^- \vert {\rm f} \rangle \langle {\rm f} \vert + \gamma^+ \vert {\rm f} \rangle \langle \vert \Omega \vert \vert - \gamma^+ \vert {\rm i} \rangle \langle {\rm i} \vert + \gamma^- \vert {\rm i} \rangle \langle 2 \vert \,,
\label{L_FiniteRTM}
\end{align}
where $\gamma^+$ and $\gamma^-$ are the forward and backward transition rates, respectively.
The backward transition is an erroneous process and can be suppressed for $\gamma^+ \gg \gamma^-$. 
However, this will be accompanied by the production of waste heat, increasing the entropy of the environment by $\ln(\gamma^+/\gamma^-)$ per step.
Computation without waste heat to the environment can be done for the undriven case, i.e., $\gamma^+ = \gamma^- = \gamma/2$.
Even under this condition, the Brownian RTM can reach the final state by chance, and we will focus on this undriven case in the following.

\begin{figure}[ht]%
\centering
\includegraphics[width=1\textwidth]{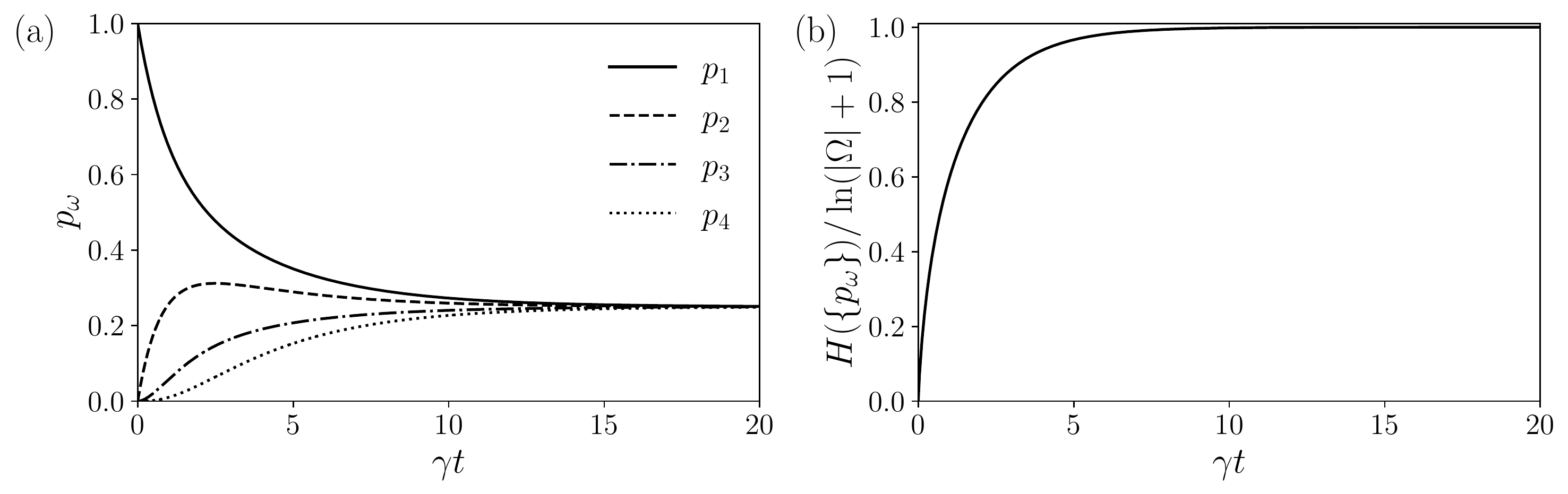}
\caption{
The time dependencies of the distribution probabilities for $\vert \Omega \vert =3$ are shown in panel (a), where the lines indicate $p_1(=p_{\rm i})$, $p_2$, $p_3(=p_{\vert \Omega \vert})$, and $p_4(=p_{\rm f})$ from top to bottom. 
The Shannon entropy is shown in panel (b).
}
\label{pH_vs_t_noreset_2_1}
\end{figure}

Figure \ref{pH_vs_t_noreset_2_1} (a) shows the time evolution of the distribution probabilities $p_\omega(t)$ of finding $\omega$th state ($\omega=1, \cdots ,\vert \Omega \vert+1$). 
Initially, the state is at $\vert 1 \rangle$, i.e., $p_1=1$ and $p_{\omega (\neq 1)}=0$. 
After a sufficiently long time $\gamma t \gg 1$, the states equilibrate and reach a uniform distribution: 
\begin{align}
p^{\rm st}_\omega=\frac{1}{\vert \Omega \vert +1} \, .
\end{align}
Accordingly, the Shannon entropy of the system,
\begin{align}
H( \{ p_\omega \} )= - \sum_{\omega} p_\omega \ln p_\omega \, ,
\end{align}
starts from zero and increases to approach its maximum value, $\ln (\vert \Omega \vert +1)$, as shown in Figure \ref{pH_vs_t_noreset_2_1} (b).

This Brownian TM behaves thermodynamically like a single Brownian particle expanding irreversibly into its configuration space~\cite{Norton2013,Strasberg2015}. 
The increase in system entropy is the thermodynamic cost of computation, since the TM must be reset to the initial zero-entropy state after computation, which requires external work $W_{\rm ext} \geq \ln (\vert \Omega \vert +1)$, as a consequence of the Landauer erasure principle. 
However, the undriven Brownian Turing machine is not useful since, at the final time, the state is unknown~\cite{Norton2013}. 
One way to ratch the `Brownian particle' at the final state $\vert {\rm f} \rangle$ is to introduce an energy trap at that state. 
Then, the change in free energy between the initial and final states after the system has equilibrated is given by $\beta \Delta F_{\rm comp} =-\ln \vert \Omega \vert - \ln (1+O_{\rm P})$, where $O_{\rm P}$ is the odds that the `Brownian particle' is at the final state and $\beta$ is the inverse temperature, as shown in Eq.(8b) of Ref.\cite{Norton2013}. 
Therefore, in the case of error-free computation where $O_{\rm P} \to \infty$, the change in free energy diverges, and consequently, the thermodynamic cost of the computation diverges.

The protocol implicitly adopted in previous works is explicitly described as follows: 
It takes into account the thermodynamic cost of computation at a given computation time (sufficiently longer than the equilibration time) with an acceptable error specified by the odds $O_{\rm P}$. 
In practice, it is more preferable to stop the computation once the Brownian particle reaches the final state. 
The duration of time to reach the final state for the first time, starting from a given initial state, is known as the first-passage time~\cite{Redner2001}. 
In the following, we will discuss the thermodynamic cost of computation in the first-passage time protocol.

For theoretical analysis, it is convenient to generalize our problem by including many resets. 
In this protocol, once the final state $\vert {\rm f} \rangle$ is reached, the system immediately resets to the initial state $\vert {\rm i} \rangle$. 
This modification changes the transition state diagram as shown in the bottom panel of Fig.~\ref{fig:rtm} (b). 
{\color{black}
The corresponding transition rate matrix is obtained by removing the final state $\vert {\rm f} \rangle$ from Eq.~(\ref{L_FiniteRTM}) and adding the unidirectional transition process for the reset as $L_{\rm RTM}'+L_{\rm reset}$:
\begin{align}
L_{\rm RTM}' =& \sum_{\omega=2}^{\vert \Omega \vert-1} \left[ - (\gamma^+ + \gamma^-) \vert \omega \rangle \langle \omega \vert + \gamma^+ \vert \omega \rangle \langle \omega-1 \vert + \gamma^- \vert \omega \rangle \langle \omega+1 \vert \right] \nonumber \\
&- \gamma^- \vert {\vert \Omega \vert} \rangle \langle {\vert \Omega \vert} \vert + \gamma^+ \vert {\vert \Omega \vert} \rangle \langle {\vert \Omega \vert}-1 \vert - \gamma^+ \vert {\rm i} \rangle \langle {\rm i} \vert + \gamma^- \vert {\rm i} \rangle \langle 2 \vert \,,
\label{L_FiniteRTM_d}
\\
L_{\rm reset} =& \gamma^+ \left( \vert {\rm i} \rangle \langle \vert \Omega \vert \vert - \vert \vert \Omega \vert \rangle \langle \vert \Omega \vert \vert \right) \, .
\label{L_reset}
\end{align}
}
The resulting transition rate matrix can be easily analyzed in the limit of {\color{black} $\gamma t \gg 1$}.

\begin{figure}[ht]%
\centering
\includegraphics[width=1\textwidth]{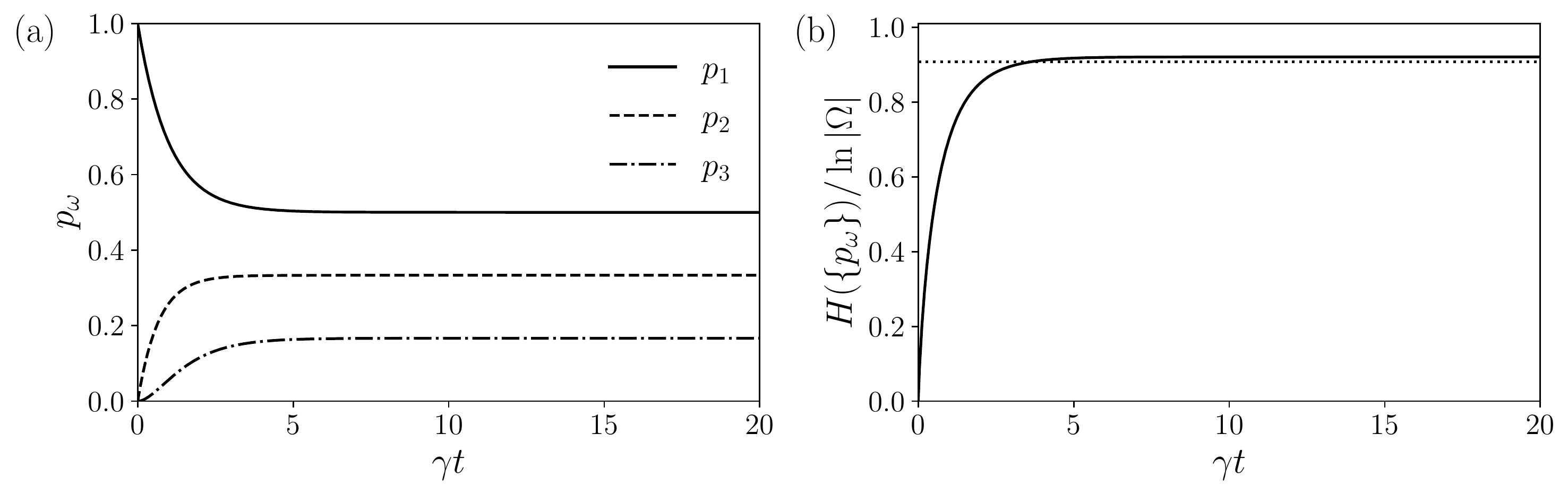}
\caption{
Time evolutions of (a) the distribution probabilities and (b) the Shannon entropy under reset boundary condition for a state space size of $\vert \Omega \vert = 3$.
The dotted line in the panel (b) corresponds to the approximation given by Eq.~(\ref{eqn:Shannon_ent_reset}), 
see also Eq.~(\ref{eqn:Shannon_ent_reset_subleading}).
}
\label{pH_vs_t_reset_2_1}
\end{figure}

Figures \ref{pH_vs_t_reset_2_1} (a) and (b) show the time dependences of the distribution probabilities and the Shannon entropy. 
As time elapses, the steady state distribution is approached (Appendix \ref{secA1}):
\begin{align}
p^{\rm st}_\omega=\frac{\vert \Omega \vert +1-\omega}{Z} \, , 
\;\;\;\; 
Z=\frac{\vert \Omega \vert (\vert \Omega \vert +1)}{2}
\, . \label{eqn:pst_reset}
\end{align}
The Shannon entropy saturates at slightly smaller than the maximum value $\ln \vert \Omega \vert$ after sufficiently long time $\gamma t \gg 1$. 
For large state space, it becomes (Appendix \ref{secA1}):
{\color{black}
\begin{align}
H( \{ p^{\rm st}_\omega \} ) 
= \ln Z - \ln H( \vert \Omega \vert )/Z
=
\ln \vert \Omega \vert -\ln2 +\frac{1}{2} -\frac{1}{2 \vert \Omega \vert} + \cdots \, , 
\label{eqn:Shannon_ent_reset}
\end{align}
where 
$H( n )= \prod_{k=1}^n k^k$
is the hyperfactorial of a positive integer $n$. 
}
The suppression from the maximum value is caused by the resets, which drive the system out of equilibrium. 
The reset process is unidirectional and requires in-depth analysis to be dealt with in a thermodynamically consistent manner, see e.g. Refs.\cite{Rahav_2014,Fuchs2016,Pal2017,Busiello2020,GuptaPRL2020,TalFriedman2020,PalPRR2021} and references therein. 
In the following, we will adopt the stochastic thermodynamics of resetting\cite{Fuchs2016}.

\section{Stochastic Thermodynamics of Resetting}\label{STR}

For a bidirectional process $(i \leftarrow j) \in E_{\rm bi}$, both the forward direction transition rate $\Gamma_{i,j} = \langle i \vert L \vert j \rangle$ and the the backward transition rate in the reverse direction $\Gamma_{j,i}$ are finite.
The flux is given by ${j}_{i,j}(t) =\Gamma_{i,j} p_{j}(t) -\Gamma_{j,i} p_{i}(t)$,
and the entropy production rates for the system and the environment are given by
\begin{align}
\langle \! \langle \sigma^{{\rm sys}}_{{\rm bi}}(t) \rangle \! \rangle = \frac{1}{2} \sum_{ (i \leftarrow j) \in E_{\rm bi}} j_{i,j}(t)  \ln \frac{ p_{j}(t) }{ p_{i}(t) }  \,   \, , \label{eqn:sig_sys_bi} \\
\langle \! \langle \sigma^{{\rm env}}_{{\rm bi}}(t) \rangle \! \rangle = \frac{1}{2} \sum_{ (i \leftarrow j) \in E_{\rm bi}}  j_{i,j}(t) \, \ln \frac{ \Gamma_{i,j} }{ \Gamma_{j,i} } \, . \label{eqn:sig_env_bi}
\end{align}

For a unidirectional process $(i \leftarrow j) \in E_{\rm uni}$, the forward transition rate $\Gamma_{i,j}$ is finite, while the backward transition rate is zero $\Gamma_{j,i}=0$.
Due to this asymmetry, the environment entropy production defined as in Eq.(\ref{eqn:sig_env_bi}) diverges. 
To address this issue, the stochastic thermodynamic of resetting~\cite{Fuchs2016} introduces a modification of the second law (Eq.~(26) of Ref.~\cite{Fuchs2016}), 
\begin{align}
\dot{H}( \{ p_j(t) \} ) + \langle \! \langle \sigma^{{\rm env}}_{{\rm bi}} \rangle \! \rangle = \langle \! \langle \sigma^{{\rm tot}}_{{\rm bi}} \rangle \! \rangle + \langle \! \langle \sigma^{{\rm sys}}_{{\rm uni}}(t) \rangle \! \rangle \geq \langle \! \langle \sigma^{{\rm sys}}_{{\rm uni}}(t) \rangle \! \rangle \, . 
\label{eqn:2ndlaw}
\end{align}
To obtain the last inequality, we utilized the fact that the total entropy production associated with the bidirectional processes is non-negative:
\begin{align}
\langle \! \langle \sigma^{{\rm tot}}_{{\rm bi}} \rangle \! \rangle =
\langle \! \langle \sigma^{{\rm sys}}_{{\rm bi}} \rangle \! \rangle +
\langle \! \langle \sigma^{{\rm env}}_{{\rm bi}} \rangle \! \rangle =
\frac{1}{2} \sum_{ (i \leftarrow j) \in E_{\rm bi}}  j_{i,j}(t) \, \ln \frac{ \Gamma_{i,j} p_j(t) }{ \Gamma_{j,i} p_i(t)}
\geq 0 \, .
\end{align}
The system entropy production rate associated with unidirectional processes 
is defined like in Eq.~(\ref{eqn:sig_sys_bi}).
It comprises entropy production rates due to absorption and insertion: 
$\langle \! \langle \sigma^{{\rm sys}}_{{\rm uni}}(t) \rangle \! \rangle = \langle \! \langle \sigma^{{\rm abs}}_{{\rm uni}}(t) \rangle \! \rangle + \langle \! \langle \sigma^{{\rm ins}}_{{\rm uni}}(t) \rangle \! \rangle$, 
where
\begin{align}
-\langle \! \langle \sigma^{{\rm abs}}_{{\rm uni}}(t) \rangle \! \rangle =& -\sum_{ (i \leftarrow j) \in E_{\rm uni}} \Gamma_{i,j} p_{j}(t) \ln p_{j}(t) \, , \label{eqn:s_abs} \\
-\langle \! \langle \sigma^{{\rm ins}}_{{\rm uni}}(t) \rangle \! \rangle =&  \sum_{ (i \leftarrow j) \in E_{\rm uni}} \Gamma_{i,j} p_{j}(t) \ln p_{i}(t) \, .  \label{eqn:s_ins} 
\end{align}
The former can be interpreted as the rate at which an external observer gains information due to the flux flowing out of state $j$, while the latter is the rate at which the oberver loses information due to the flux flowing into state $i$. 
The negative of the resetting entropy production, $-\langle \! \langle \sigma^{{\rm sys}}_{{\rm uni}}(t) \rangle \! \rangle$, provides the lower bound of the amount of external work required for resetting, see Eq.~(30) of Ref.~\cite{Fuchs2016}.  
If there is no unidirectional process, the left-hand side of Eq.~(\ref{eqn:2ndlaw}) is the ordinary entropy production rate. 
Note that when unidirectional processes are present, the left-hand side of Eq.~(\ref{eqn:2ndlaw}) can be negative.

Next, we consider stochastic thermodynamics at the single-trajectory level~\cite{Seifert2012,Pal2017,Busiello2020}. 
Suppose that the system undergoes a transition from state $\omega_k$ to $\omega_{k+1}$ at time $t=t_{k+1}$. 
The stochastic trajectory consists of a sequence of $N+1$ times, including the initial time $t_0=0$ and $N$ successive times at which transitions occur:
$t^{N+1}=t_0 t_1 \cdots t_N$,
and the $N+1$ successive states:
$\omega^{N+1}=\omega_0 \omega_1 \cdots \omega_N$.
The final time $t_N=\tau$ is the first-passage time. The initial and final states are given by $\omega_0={\rm i}$ and $\omega_N={\rm f}$, respectively. 
The other states should not include the final states, i.e., $\omega_n \neq {\rm f}$ for $n=1,\cdots, N-1$.

The stochastic entropy~\cite{Seifert2012,Busiello2020} is a key quantity defined for each trajectory in stochastic thermodynamics. 
Its time derivative is given by
\begin{align}
\dot{\Sigma}^{\rm sys}(t \vert t^{N+1},\omega^{N+1})= - \partial_t \ln p_{\omega}(t) \vert_{\omega = \omega(t)} - \sum_{n=1}^N \delta(t -t_n) \ln \frac{p_{\omega_{n}}(t_n)}{p_{\omega_{n-1}}(t_n)}  \, .
\end{align}
where $p_{\omega}(t)$ is the distribution probability of finding the system in state $\omega$ at time $t$. 
The sum in the second term is performed over all transition times $t_j$ in the trajectory, and measures the entropy production associated with each transition.
The change in stochastic entropy for a given trajectory is obtained by integrating the time derivative over the duration of the trajectory from $t=0$ to $t=\tau$, which gives, 
\begin{align}
{\Sigma}^{\rm sys}(t^{N+1},\omega^{N+1}) 
= 
\ln \frac{p_{\omega_{0}} (t_{0})}{p_{\omega_{N}} (t_{N})}  
=
\Sigma^{\rm sys}_{\rm bi} \left( \omega^{N+1},t^{N+1} \right)
+
\Sigma^{\rm sys}_{\rm uni} \left( \omega^{N+1},t^{N+1} \right)
\, ,
\label{eqn:path_ent}
\end{align}
where the first term on the right-hand side is the contribution from the bidirectional transitions and the second term is the contribution from the unidirectional transitions.
The latter is given by
\begin{align}
\Sigma^{\rm sys}_{\rm uni} \left( \omega^{N+1},t^{N+1} \right) = \sum_{n=1}^{N} \ln \frac{ p_{\omega_{n-1}} (t_{n}) }{ p_{\omega_{n}} (t_{n}) } \phi_{E_{\rm uni}}( (\omega_{n} \leftarrow \omega_{n-1})  ) \, ,
\label{eqn:path_ent_pro_uni} 
\end{align}
where the summation is performed only for the unidirectional transitions, and the indicator function $\phi_E(x)$ is equal to 1 if $x$ belongs to the set $E$ and zero otherwise.

In standard numerical methods, such as the Gillespie algorithm~\cite{Gillespie1976,Gillespie1977}, trajectories are constructed using random numbers. 
This makes it convenient to calculate quantities defined at a single-trajectory level.
However, Eq.~(\ref{eqn:path_ent_pro_uni}) includes the distribution probability $p_{\omega}(t)$, which cannot be determined from a single trajectory. 
Therefore, naively, we must calculate the distribution probability $p_\omega(t)$ at all times in advance by solving the master equation with a given initial distribution probability. 

To circumvent this technical problem, we adopt an alternative approach in Ref.\cite{utsumi2022computation}, in which the distribution probability is approximated with the empirical sojourn time given by, 
\begin{align}
\tau_{\omega} \left( t^{N+1} \right) =& \sum_{n=1}^{N} (t_{n}-t_{n-1}) \delta_{\omega_{n-1}, \, \omega} \, , \label{eq:sojourn}
\end{align}
as $p_{\omega_{n}}/p_{\omega_{n-1}} \approx \tau_{\omega_{n}}/\tau_{\omega_{n-1}}$ in Eq.~(\ref{eqn:path_ent_pro_uni}). 
The empirical sojourn time is determined only from a single trajectory, thereby avoiding the need of $p_\omega(t)$ at all times. 
In the next section, we will discuss the validity of this approximation.

\section{Results and Discussion}\label{RD}

\begin{figure}[ht]%
\centering
\includegraphics[width=1\textwidth]{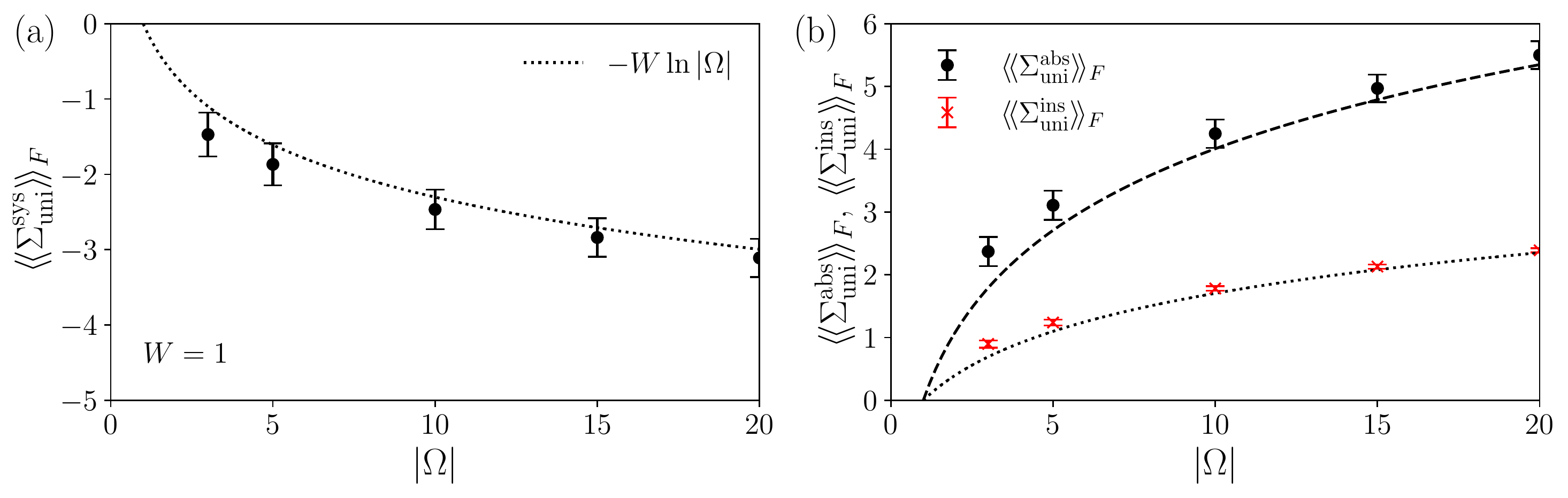}
\caption{
Dependences of (a) resetting entropy production and (b) entropy productions due to absorption and insertion on the state-space size for $W=1$. 
In panel (b), the dashed and dotted lines represent the analytic results given by Eqs.~(\ref{eqn:s_abs_ins_RTM}).
The number of samples is $10^5$, and error bars indicate the standard deviation.
}
\label{res_ent_vs_size_1_}
\end{figure}

Figure \ref{res_ent_vs_size_1_}(a) shows the resetting entropy calculated using the Gillespie algorithm. 
The time-dependent distribution probability is calculated by solving the master equation, as shown in panel (a) of Fig.~\ref{pH_vs_t_reset_2_1}. 
The resetting entropy is negative, indicating that the reset decreases the system entropy. The curve is well fitted by $-W \ln \vert \Omega \vert$ (dotted line), which is consistent with previous theories based on the Landauer principle~\cite{Norton2013,Strasberg2015}.

The interpretation in terms of the stochastic thermodynamics of resetting is as follows: 
In the limit of many resets $W \gg 1$, the entropy productions can be estimated using the steady-state distribution probability. 
The rates of entropy productions due to absorption and insertion are 
$\langle \! \langle \sigma^{{\rm abs}}_{{\rm uni}} \rangle \! \rangle = (\gamma/2) p^{\rm st}_{\vert \Omega \vert} \ln { p^{\rm st}_{\vert \Omega \vert} }$
and
$\langle \! \langle \sigma^{{\rm ins}}_{{\rm uni}} \rangle \! \rangle = - (\gamma/2) p^{\rm st}_{\vert \Omega \vert} \ln { p^{\rm st}_{1} }$. 
The entropy productions in the first-passage time protocol are the product of their rates and the average first passage time,
$\langle \! \langle \tau \rangle \! \rangle_F = W/\langle \! \langle w \rangle \! \rangle$, where $W$ is the number of resets and $\langle \! \langle w \rangle \! \rangle = (\gamma/2) p^{\rm st}_{\vert \Omega \vert}$ is the averate reset curent. 
Therefore, the absorption and insertion entropies are given by
\begin{align}
\langle \! \langle \Sigma^{{\rm abs}}_{{\rm uni}} \rangle \! \rangle_F = - W \ln \frac{ \vert \Omega \vert ( \vert \Omega \vert+1 )}{2}  \, , \;\;\;\;
\langle \! \langle \Sigma^{{\rm ins}}_{{\rm uni}} \rangle \! \rangle_F = W \ln \frac{\vert \Omega \vert+1}{2}  \, . \label{eqn:s_abs_ins_RTM} 
\end{align}

Figure \ref{res_ent_vs_size_1_} (b) shows the dependence of absorption and insertion entropies on the state-space size, where the dashed and dotted lines represent the analytic results in Eq.~(\ref{eqn:s_abs_ins_RTM}). 
For large state-space size, the numerical results are well fitted by the analytical expressions even for $W=1$. 
As the state space size increases, the absoption reduced the system entropy by 
$\langle \! \langle \Sigma^{{\rm abs}}_{{\rm uni}} \rangle \! \rangle_F \approx - 2 W \ln \vert \Omega \vert$, 
while the insertion increases the system entropy by 
$\langle \! \langle \Sigma^{{\rm ins}}_{{\rm uni}} \rangle \! \rangle_F \approx W \ln \vert \Omega \vert$. 
The sum of the two contributions is the reset entropy production, providing the prescribed result, 
$\langle \! \langle \Sigma^{{\rm sys}}_{{\rm uni}} \rangle \! \rangle_F =
\langle \! \langle \Sigma^{{\rm abs}}_{{\rm uni}} \rangle \! \rangle_F +
\langle \! \langle \Sigma^{{\rm ins}}_{{\rm uni}} \rangle \! \rangle_F = -W \ln \vert \Omega \vert$.

\begin{figure}[ht]%
\centering
\includegraphics[width=1\textwidth]{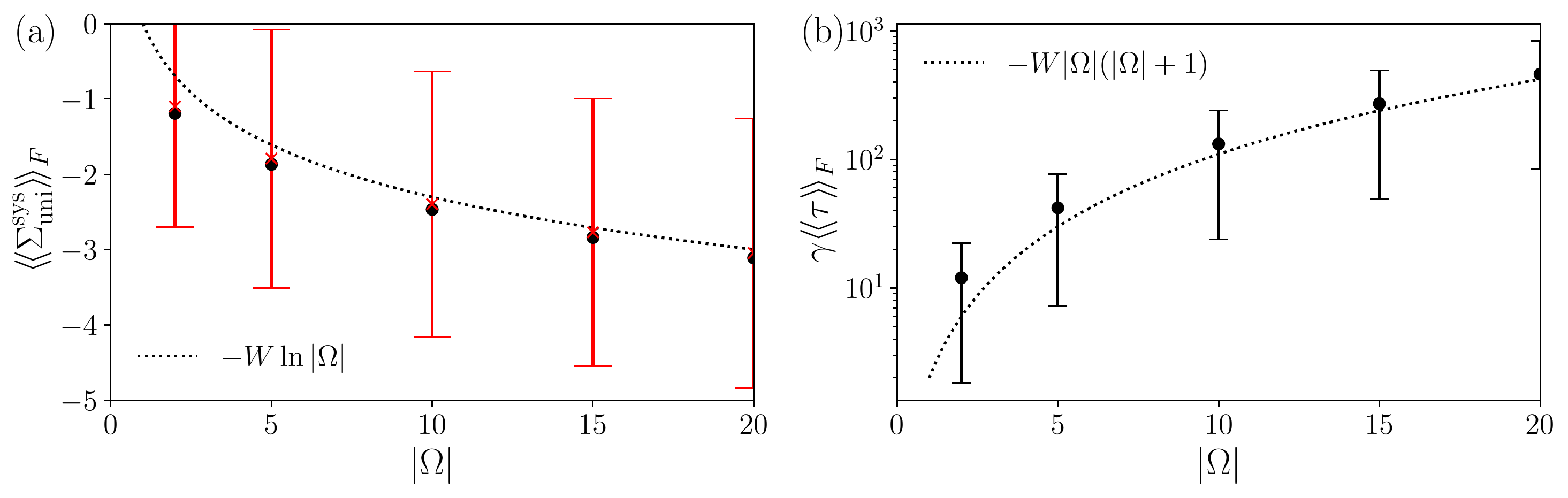}
\caption{
The state-space size dependences of (a) the resetting entropy production (crosses with error bars) and (b) the computation time for $W=1$. 
In panel (a), solid circles indicate the results in panel (a) of Fig.~\ref{res_ent_vs_size_1_}.
}
\label{res_ent_vs_size_soj1_}
\end{figure}

Figure~\ref{res_ent_vs_size_soj1_} (a) shows the resetting entropy production obtained by approximating the ratio of distribution probabilities by the ratio of sojourn times, Eq.(\ref{eq:sojourn}). 
The average values (crosses) well reproduce the results shown in panel (a) of Fig.\ref{res_ent_vs_size_1_} (solid circles). 
However, it overestimates the width of the distribution indicated by error bars. 
This is expected since the empirical sojourn time is a fluctuating quantity and is distributed accordingly~\cite{YU2007}. 
Despite this limitation, this approximation would be useful when the system size is large~\cite{utsumi2022computation}. 

The panel (b) of Fig.~\ref{res_ent_vs_size_soj1_} shows the state-space size dependence of the computation time, which increases as $\langle \! \langle \tau \rangle \! \rangle_F \approx W \vert \Omega \vert^2$. This behavior is reasonable since in an infinite one-dimensional chain, the variance of the location of the `Brownian particle' grows as $ \langle \! \langle \delta \omega(\tau)^2 \rangle \! \rangle \sim \tau$, and thus the time to reach the final state roughly estimated from $\sqrt{ \langle \! \langle \delta \omega(\tau)^2 \rangle \! \rangle } \approx \vert \Omega \vert$, is $\tau \sim \vert \Omega \vert^2$.

\section{Conclusion}\label{conclusion}

We discuss the thermodynamic cost of a logically reversible Turing machine operating in the first-passage time protocol. 
The thermodynamic cost, which is the negative of the resetting entropy production, exhibits a logarithmic dependence on the size of the state space. 
By using the Gillespie algorithm, we numerically calculate the stochastic entropy associated with unidirectional resetting processes. 
For this purpose, we calculate the time-dependent distribution probability in advance by solving the master equation.
We analyze an approximation that replaces the time-dependent distribution probability with the empirical sojourn time, which can be computed from a single trajectory.
While this approximation reproduces the average reset entropy reasonably well, it overestimates its fluctuations. 

\backmatter

\bmhead{Acknowledgments}
This work was supported by JSPS KAKENHI Grants No. 18KK0385, No. 20H01827, and No. 20H05666, and JST, CREST Grant Number JPMJCR20C1, Japan. 

\bmhead{Data Availability Statement}
The data that support the findings of this study are available from the corresponding author upon reasonable request.

\begin{appendices}

\section{Derivation of Eqs.~(\ref{eqn:pst_reset}) and (\ref{eqn:Shannon_ent_reset}) }\label{secA1}

{\color{black}
One can check $\vert p^{\rm st} \rangle = \sum_{\omega=1}^{\vert \Omega \vert} \vert \omega \rangle p^{\rm st}_\omega $
satisfies 
$ \langle \omega \vert \left( L_{\rm RTM}'+L_{\rm reset} \right) \vert p^{\rm st} \rangle=0$,
which is 
\begin{align}
(\gamma/2) (- 2 p^{\rm st}_\omega + p^{\rm st}_{\omega+1} + p^{\rm st}_{\omega-1}) &=0 \, , 
\;\;\;\; (\omega = 2,\cdots, \vert \Omega \vert-1) \, , \label{eqn:bulk}
\\
(\gamma/2) (-p^{\rm st}_1 + p^{\rm st}_{2} + p^{\rm st}_{\vert \Omega \vert}) &=0 \, ,
\;\;\;\; (\omega = 1) \, , \label{eqn:edge1} \\
(\gamma/2) (-2 p^{\rm st}_{\vert \Omega \vert} + p^{\rm st}_{\vert \Omega \vert-1}) &=0 \, , 
\;\;\;\; (\omega = \vert \Omega \vert) \, . \label{eqn:edge2}
\end{align}
Equation~(\ref{eqn:bulk}) represents the Laplace equation $(\gamma/2) \Delta p_\omega^{\rm st}=0$, 
which has a solution of the form $p_\omega=a+b \omega$, where $a$ and $b$ are constants. 
Substituting this solution into Eqs.~(\ref{eqn:edge1}) and (\ref{eqn:edge2}), we obtain $a+b(\vert \Omega \vert + 1)=0$. Combining this result with the normalization condition $\sum_{\omega=1}^{\vert \Omega \vert} p_\omega^{\rm st}=1$, we arrive at Eq. (\ref{eqn:pst_reset}).

The first equality in Eq. (\ref{eqn:Shannon_ent_reset}) can be obtained through a straightforward calculation or by utilizing 
$H( \{ p_j^{\rm st} \} )=\lim_{q \to 1} H_q( \{ p_j^{\rm st} \} )$,
where
\begin{align}
H_q( \{ p_j^{\rm st} \} ) = \frac{\ln \sum_{\omega=1}^{\vert \Omega \vert} (p^{\rm st}_\omega)^q }{1-q} =
\frac{ \ln H_{\vert \Omega \vert}^{(-q)} Z^{-q}}{1-q} \, ,
\end{align}
represents the R\'enyi entropy, and $H_n^{(r)} = \sum_{k=1}^n k^{-r}$ denotes the generalized harmonic number of order $r$ for $n$.

By using $H( n ) \approx A n^{n(n+1)/2 + 1/12} e^{-n^2/4}$ for $n \gg 1$, where $A$ is the Glaisher-Kinkelin constant, 
Eq.~(\ref{eqn:Shannon_ent_reset}) is expanded in $1/\vert { \Omega } \vert$ as, 
\begin{align}
H( \{ p^{\rm st}_\omega \} )\approx \ln \vert \Omega \vert -\ln2 +\frac{1}{2} -\frac{1}{2 \vert \Omega \vert} 
- \frac{\ln \vert \Omega \vert +12 \ln A}{6 \vert \Omega \vert^2} \, .
\label{eqn:Shannon_ent_reset_subleading}
\end{align}
}




\end{appendices}


\bibliography{sn-bibliography_yu}


\begin{thebibliography}{28}
\ifx \bisbn   \undefined \def \bisbn  #1{ISBN #1}\fi
\ifx \binits  \undefined \def \binits#1{#1}\fi
\ifx \bauthor  \undefined \def \bauthor#1{#1}\fi
\ifx \batitle  \undefined \def \batitle#1{#1}\fi
\ifx \bjtitle  \undefined \def \bjtitle#1{#1}\fi
\ifx \bvolume  \undefined \def \bvolume#1{\textbf{#1}}\fi
\ifx \byear  \undefined \def \byear#1{#1}\fi
\ifx \bissue  \undefined \def \bissue#1{#1}\fi
\ifx \bfpage  \undefined \def \bfpage#1{#1}\fi
\ifx \blpage  \undefined \def \blpage #1{#1}\fi
\ifx \burl  \undefined \def \burl#1{\textsf{#1}}\fi
\ifx \doiurl  \undefined \def \doiurl#1{\url{https://doi.org/#1}}\fi
\ifx \betal  \undefined \def \betal{\textit{et al.}}\fi
\ifx \binstitute  \undefined \def \binstitute#1{#1}\fi
\ifx \binstitutionaled  \undefined \def \binstitutionaled#1{#1}\fi
\ifx \bctitle  \undefined \def \bctitle#1{#1}\fi
\ifx \beditor  \undefined \def \beditor#1{#1}\fi
\ifx \bpublisher  \undefined \def \bpublisher#1{#1}\fi
\ifx \bbtitle  \undefined \def \bbtitle#1{#1}\fi
\ifx \bedition  \undefined \def \bedition#1{#1}\fi
\ifx \bseriesno  \undefined \def \bseriesno#1{#1}\fi
\ifx \blocation  \undefined \def \blocation#1{#1}\fi
\ifx \bsertitle  \undefined \def \bsertitle#1{#1}\fi
\ifx \bsnm \undefined \def \bsnm#1{#1}\fi
\ifx \bsuffix \undefined \def \bsuffix#1{#1}\fi
\ifx \bparticle \undefined \def \bparticle#1{#1}\fi
\ifx \barticle \undefined \def \barticle#1{#1}\fi
\bibcommenthead
\ifx \bconfdate \undefined \def \bconfdate #1{#1}\fi
\ifx \botherref \undefined \def \botherref #1{#1}\fi
\ifx \url \undefined \def \url#1{\textsf{#1}}\fi
\ifx \bchapter \undefined \def \bchapter#1{#1}\fi
\ifx \bbook \undefined \def \bbook#1{#1}\fi
\ifx \bcomment \undefined \def \bcomment#1{#1}\fi
\ifx \oauthor \undefined \def \oauthor#1{#1}\fi
\ifx \citeauthoryear \undefined \def \citeauthoryear#1{#1}\fi
\ifx \endbibitem  \undefined \def \endbibitem {}\fi
\ifx \bconflocation  \undefined \def \bconflocation#1{#1}\fi
\ifx \arxivurl  \undefined \def \arxivurl#1{\textsf{#1}}\fi
\csname PreBibitemsHook\endcsname

\bibitem{Landauer1961}
\begin{barticle}
\bauthor{\bsnm{Landauer}, \binits{R.}}:
\batitle{Irreversibility and heat generation in the computing process}.
\bjtitle{IBM Journal of Research and Development}
\bvolume{5}(\bissue{3}),
\bfpage{183}--\blpage{191}
(\byear{1961}).
\doiurl{10.1147/rd.53.0183}
\end{barticle}
\endbibitem

\bibitem{Bennett1982}
\begin{barticle}
\bauthor{\bsnm{Bennett}, \binits{C.H.}}:
\batitle{The thermodynamics of computation—a review}.
\bjtitle{International Journal of Theoretical Physics}
\bvolume{21}(\bissue{12}),
\bfpage{905}--\blpage{940}
(\byear{1982}).
\doiurl{10.1007/BF02084158}
\end{barticle}
\endbibitem

\bibitem{Bennett1985}
\begin{barticle}
\bauthor{\bsnm{Bennett}, \binits{C.H.}},
\bauthor{\bsnm{Landauer}, \binits{R.}}:
\batitle{The fundamental physical limits of computation}.
\bjtitle{Scientific American}
\bvolume{253},
\bfpage{48}--\blpage{56}
(\byear{1985})
\end{barticle}
\endbibitem

\bibitem{Sagawa2019}
\begin{botherref}
\oauthor{\bsnm{Sagawa}, \binits{T.}}:
Second law, entropy production, and reversibility in thermodynamics of information.
Energy Limits in Computation: A Review of Landauer's Principle, Theory and Experiments
(2019)
\end{botherref}
\endbibitem

\bibitem{Wolpert2019}
\begin{barticle}
\bauthor{\bsnm{Wolpert}, \binits{D.H.}}:
\batitle{The stochastic thermodynamics of computation}.
\bjtitle{Journal of Physics A: Mathematical and Theoretical}
\bvolume{52}(\bissue{19}),
\bfpage{193001}
(\byear{2019}).
\doiurl{10.1088/1751-8121/ab0850}
\end{barticle}
\endbibitem

\bibitem{Kolchinsky2020}
\begin{barticle}
\bauthor{\bsnm{Kolchinsky}, \binits{A.}},
\bauthor{\bsnm{Wolpert}, \binits{D.H.}}:
\batitle{Thermodynamic costs of turing machines}.
\bjtitle{Phys. Rev. Res.}
\bvolume{2},
\bfpage{033312}
(\byear{2020}).
\doiurl{10.1103/PhysRevResearch.2.033312}
\end{barticle}
\endbibitem

\bibitem{Ito2013}
\begin{barticle}
\bauthor{\bsnm{Ito}, \binits{S.}},
\bauthor{\bsnm{Sagawa}, \binits{T.}}:
\batitle{Information thermodynamics on causal networks}.
\bjtitle{Phys. Rev. Lett.}
\bvolume{111},
\bfpage{180603}
(\byear{2013}).
\doiurl{10.1103/PhysRevLett.111.180603}
\end{barticle}
\endbibitem

\bibitem{Seifert2012}
\begin{barticle}
\bauthor{\bsnm{Seifert}, \binits{U.}}:
\batitle{Stochastic thermodynamics, fluctuation theorems and molecular machines}.
\bjtitle{Reports on Progress in Physics}
\bvolume{75}(\bissue{12}),
\bfpage{126001}
(\byear{2012}).
\doiurl{10.1088/0034-4885/75/12/126001}
\end{barticle}
\endbibitem

\bibitem{VandenBroeck2013}
\begin{bchapter}
\bauthor{\bparticle{Van~den} \bsnm{Broeck}, \binits{C.}}:
\bctitle{Stochastic thermodynamics: A brief introduction}.
In: \beditor{\bsnm{Bechinger}, \binits{C.}},
\beditor{\bsnm{Sciortino}, \binits{F.}},
\beditor{\bsnm{Ziherl}, \binits{P.}} (eds.)
\bbtitle{Proceedings of the International School of Physics "Enrico Fermi" Course CLXXXIV "Physics of Complex Colloids"}.
\bpublisher{IOS},
\blocation{Amsterdam}
(\byear{2013})
\end{bchapter}
\endbibitem

\bibitem{VandenBroeck2015}
\begin{barticle}
\bauthor{\bsnm{{Van den Broeck}}, \binits{C.}},
\bauthor{\bsnm{Esposito}, \binits{M.}}:
\batitle{Ensemble and trajectory thermodynamics: A brief introduction}.
\bjtitle{Physica A: Statistical Mechanics and its Applications}
\bvolume{418},
\bfpage{6}--\blpage{16}
(\byear{2015}).
\doiurl{10.1016/j.physa.2014.04.035}.
\bcomment{Proceedings of the 13th International Summer School on Fundamental Problems in Statistical Physics}
\end{barticle}
\endbibitem

\bibitem{Klages2012}
\begin{bbook}
\beditor{\bsnm{Klages}, \binits{R.}},
\beditor{\bsnm{Just}, \binits{W.}},
\beditor{\bsnm{Jarzynski}, \binits{C.}} (eds.):
\bbtitle{Nonequilibrium Statistical Physics of Small Systems: Fluctuation Relations and Beyond}.
\bpublisher{Wiley-VCH},
\blocation{Weinheim}
(\byear{2012})
\end{bbook}
\endbibitem

\bibitem{Pekola2015}
\begin{barticle}
\bauthor{\bsnm{Pekola}, \binits{J.P.}}:
\batitle{Towards quantum thermodynamics in electronic circuits}.
\bjtitle{Nature Physics}
\bvolume{11}(\bissue{2}),
\bfpage{118}--\blpage{123}
(\byear{2015}).
\doiurl{10.1038/nphys3169}
\end{barticle}
\endbibitem

\bibitem{Norton2013}
\begin{barticle}
\bauthor{\bsnm{Norton}, \binits{J.D.}}:
\batitle{Brownian computation is thermodynamically irreversible}.
\bjtitle{Foundations of Physics}
\bvolume{43}(\bissue{11}),
\bfpage{1}--\blpage{27}
(\byear{2013}).
\doiurl{10.1007/s10701-013-9753-1}
\end{barticle}
\endbibitem

\bibitem{Strasberg2015}
\begin{barticle}
\bauthor{\bsnm{Strasberg}, \binits{P.}},
\bauthor{\bsnm{Cerrillo}, \binits{J.}},
\bauthor{\bsnm{Schaller}, \binits{G.}},
\bauthor{\bsnm{Brandes}, \binits{T.}}:
\batitle{Thermodynamics of stochastic turing machines}.
\bjtitle{Phys. Rev. E}
\bvolume{92},
\bfpage{042104}
(\byear{2015}).
\doiurl{10.1103/PhysRevE.92.042104}
\end{barticle}
\endbibitem

\bibitem{Fuchs2016}
\begin{barticle}
\bauthor{\bsnm{Fuchs}, \binits{J.}},
\bauthor{\bsnm{Goldt}, \binits{S.}},
\bauthor{\bsnm{Seifert}, \binits{U.}}:
\batitle{Stochastic thermodynamics of resetting}.
\bjtitle{Europhysics Letters}
\bvolume{113}(\bissue{6}),
\bfpage{60009}
(\byear{2016}).
\doiurl{10.1209/0295-5075/113/60009}
\end{barticle}
\endbibitem

\bibitem{utsumi2022computation}
\begin{botherref}
\oauthor{\bsnm{Utsumi}, \binits{Y.}},
\oauthor{\bsnm{Ito}, \binits{Y.}},
\oauthor{\bsnm{Golubev}, \binits{D.}},
\oauthor{\bsnm{Peper}, \binits{F.}}:
Computation time and thermodynamic uncertainty relation of brownian circuits
(2022)
{\href{https://arxiv.org/abs/2205.10735}{{arXiv:2205.10735}}}
{[cond-mat.stat-mech]}
\end{botherref}
\endbibitem

\bibitem{toledomarin2022passage}
\begin{botherref}
\oauthor{\bsnm{Toledo-Marin}, \binits{J.Q.}},
\oauthor{\bsnm{Boyer}, \binits{D.}}:
First passage time and information of a one-dimensional brownian particle with stochastic resetting to random positions
(2022)
{\href{https://arxiv.org/abs/2206.14387}{{arXiv:2206.14387}}}
{[cond-mat.soft]}
\end{botherref}
\endbibitem

\bibitem{Bennett1973}
\begin{barticle}
\bauthor{\bsnm{Bennett}, \binits{C.H.}}:
\batitle{Logical reversibility of computation}.
\bjtitle{IBM Journal of Research and Development}
\bvolume{17}(\bissue{6}),
\bfpage{525}--\blpage{532}
(\byear{1973}).
\doiurl{10.1147/rd.176.0525}
\end{barticle}
\endbibitem

\bibitem{Redner2001}
\begin{bbook}
\bauthor{\bsnm{Redner}, \binits{S.}}:
\bbtitle{A Guide to First-Passage Processes}.
\bpublisher{Cambridge University Press},
\blocation{Cambridge}
(\byear{2001})
\end{bbook}
\endbibitem

\bibitem{Rahav_2014}
\begin{barticle}
\bauthor{\bsnm{Rahav}, \binits{S.}},
\bauthor{\bsnm{Harbola}, \binits{U.}}:
\batitle{An integral fluctuation theorem for systems with unidirectional transitions}.
\bjtitle{Journal of Statistical Mechanics: Theory and Experiment}
\bvolume{2014}(\bissue{10}),
\bfpage{10044}
(\byear{2014}).
\doiurl{10.1088/1742-5468/2014/10/P10044}
\end{barticle}
\endbibitem

\bibitem{Pal2017}
\begin{barticle}
\bauthor{\bsnm{Pal}, \binits{A.}},
\bauthor{\bsnm{Rahav}, \binits{S.}}:
\batitle{Integral fluctuation theorems for stochastic resetting systems}.
\bjtitle{Phys. Rev. E}
\bvolume{96},
\bfpage{062135}
(\byear{2017}).
\doiurl{10.1103/PhysRevE.96.062135}
\end{barticle}
\endbibitem

\bibitem{Busiello2020}
\begin{barticle}
\bauthor{\bsnm{Busiello}, \binits{D.M.}},
\bauthor{\bsnm{Gupta}, \binits{D.}},
\bauthor{\bsnm{Maritan}, \binits{A.}}:
\batitle{Entropy production in systems with unidirectional transitions}.
\bjtitle{Phys. Rev. Res.}
\bvolume{2},
\bfpage{023011}
(\byear{2020}).
\doiurl{10.1103/PhysRevResearch.2.023011}
\end{barticle}
\endbibitem

\bibitem{GuptaPRL2020}
\begin{barticle}
\bauthor{\bsnm{Gupta}, \binits{D.}},
\bauthor{\bsnm{Plata}, \binits{C.A.}},
\bauthor{\bsnm{Pal}, \binits{A.}}:
\batitle{Work fluctuations and jarzynski equality in stochastic resetting}.
\bjtitle{Phys. Rev. Lett.}
\bvolume{124},
\bfpage{110608}
(\byear{2020}).
\doiurl{10.1103/PhysRevLett.124.110608}
\end{barticle}
\endbibitem

\bibitem{TalFriedman2020}
\begin{barticle}
\bauthor{\bsnm{Tal-Friedman}, \binits{O.}},
\bauthor{\bsnm{Pal}, \binits{A.}},
\bauthor{\bsnm{Sekhon}, \binits{A.}},
\bauthor{\bsnm{Reuveni}, \binits{S.}},
\bauthor{\bsnm{Roichman}, \binits{Y.}}:
\batitle{Experimental realization of diffusion with stochastic resetting}.
\bjtitle{The Journal of Physical Chemistry Letters}
\bvolume{11}(\bissue{17}),
\bfpage{7350}--\blpage{7355}
(\byear{2020})
{\href{https://arxiv.org/abs/https://doi.org/10.1021/acs.jpclett.0c02122}{{https://doi.org/10.1021/acs.jpclett.0c02122}}}.
\doiurl{10.1021/acs.jpclett.0c02122}.
\bcomment{PMID: 32787296}
\end{barticle}
\endbibitem

\bibitem{PalPRR2021}
\begin{barticle}
\bauthor{\bsnm{Pal}, \binits{A.}},
\bauthor{\bsnm{Reuveni}, \binits{S.}},
\bauthor{\bsnm{Rahav}, \binits{S.}}:
\batitle{Thermodynamic uncertainty relation for systems with unidirectional transitions}.
\bjtitle{Phys. Rev. Res.}
\bvolume{3},
\bfpage{013273}
(\byear{2021}).
\doiurl{10.1103/PhysRevResearch.3.013273}
\end{barticle}
\endbibitem

\bibitem{Gillespie1976}
\begin{barticle}
\bauthor{\bsnm{Gillespie}, \binits{D.T.}}:
\batitle{A general method for numerically simulating the stochastic time evolution of coupled chemical reactions}.
\bjtitle{Journal of Computational Physics}
\bvolume{22}(\bissue{4}),
\bfpage{403}--\blpage{434}
(\byear{1976}).
\doiurl{10.1016/0021-9991(76)90041-3}
\end{barticle}
\endbibitem

\bibitem{Gillespie1977}
\begin{barticle}
\bauthor{\bsnm{Gillespie}, \binits{D.T.}}:
\batitle{Exact stochastic simulation of coupled chemical reactions}.
\bjtitle{The Journal of Physical Chemistry}
\bvolume{81}(\bissue{25}),
\bfpage{2340}--\blpage{2361}
(\byear{1977})
{\href{https://arxiv.org/abs/https://doi.org/10.1021/j100540a008}{{https://doi.org/10.1021/j100540a008}}}.
\doiurl{10.1021/j100540a008}
\end{barticle}
\endbibitem

\bibitem{YU2007}
\begin{barticle}
\bauthor{\bsnm{Utsumi}, \binits{Y.}}:
\batitle{Full counting statistics for the number of electrons in a quantum dot}.
\bjtitle{Phys. Rev. B}
\bvolume{75},
\bfpage{035333}
(\byear{2007}).
\doiurl{10.1103/PhysRevB.75.035333}
\end{barticle}
\endbibitem

\end{thebibliography}


\end{document}